# INTRODUCING ENRICHED CONCRETE SYNTAX TREE

*Gordana Rakić, Zoran Budimac*
Dept. of Mathematics and Informatics, Faculty of Sciences, University of Novi Sad
Trg Dositeja Obradovića 4, 21000 Novi Sad, Serbia
+381 21 4852877, +381 21 458888
goca@dmi.uns.ac.rs, zjb@dmi.uns.ac.rs

**ABSTRACT**

**In our earlier research [9] an area of consistent and systematic application of software metrics was explored. Strong dependency of applicability of software metrics on input programming language was recognized as one of the main weaknesses in this field. Introducing enriched Concrete Syntax Tree (eCST) for internal and intermediate representation of the source code resulted with step forward over this weakness. In this paper we explain innovation made by introducing eCST and provide idea for broader applicability of eCST in some other fields of software engineering.**

## 1 INTRODUCTION

We introduce the new type of Syntax Trees to be used as intermediate representation of the source code. We call this type of tree "enriched Concrete Syntax Tree (eCST), because it is enriched by so-called "universal" nodes. Universal nodes are common for all programming languages and add additional value to these trees. In this way the tree becomes broadly usable for different algorithms of software metrics and independent on input programming language. Besides original application field in building software metrics tools and other static analyzers, eCSTs can be further manipulated and transformed and thus applied in other fields as well (e.g. software testing and source code translation). In this way eCST can be used for numerous purposes related to development, maintenance and analysis of software systems. The major benefit of usage of eCST is its universality and independency on programming language.

## 2 BACKGROUND

Syntax trees are usually secondary product of compiler and parser generators. These generators usually have embedded mechanisms to generate syntax trees as internal structures. Additionally, these mechanisms can be extended with mechanism for enrichment of syntax trees with additional information about language or input source code. This opportunity is our key instrument.

Parser generators take the language grammar as its input and return parser for that language as an output. This grammar is provided in some default form (e.g. EBNF – Extended Backus-Naur form) or in some generator-specific syntax. In this project parser generator is used to generate scanner and parser with embedded rules and functions for generating trees and for managing the content of its nodes. In fact, we use it to generate parser that will produce eCST including insertion of universal nodes.

The main idea is to add corresponding universal node as a parent of sub-tree that represents specific element in a source code (e.g. COMPILATION_UNIT, FUNCTION_CALL, BRANCH_STATEMENT, LOOP_STATEMENT, etc.). Key characteristic of these nodes is that these are equivalent in all programming languages.

## 3 RELATED WORK

Syntax trees, abstract or concrete, are broadly used in numerous fields of software engineering. Abstract Syntax Tree (AST) is used as representation of source code and model.

Baxter [1] and Ducasse [2] use abstract syntax trees for representation of the source code for duplicated code analysis. Those trees have some additional features designed for easier implementation of the algorithm for comparison. Koschke et al. [5] propose similar but fresh idea for code clone detection using abstract syntax suffix trees

In [3] the role of AST as representation of model in Model Driven Engineering is described. ASTs were also used for monitoring of changes (Neamtiu et al. [6]) in the change analysis of code written in programming language C.

Even if the construction of AST is language independent; the content of these trees is always strongly related to language syntax. That can be clearly concluded from all papers related to usage of AST referred in this article.

In [9] a detailed motivation for initiating the research in proposed direction is described. It is related to problems in application of software metrics and early work on introducing eCST in that particular field. In [9, 10] is described eCST, its construction and its role in development of SMIILE (Software Metrics - Independent of Input LanguagE) tool, as well.

Additionally, we can propose [7] as an introduction to automatic building of syntax trees by generated language parser. It also provides mechanism for adding universal nodes into tree that is to be generated.

## 4 INTRODUCING eCST

Related research shows that there is no fully consistent support for software development and maintenance. All tools used for these purposes have some limitations, e.g. limited programming language support, weak and inconsistent usage of metrics and/or testing techniques, etc.

In the field of software evolution, which enforces techniques such are advising, recommending and automating of refactoring and reengineering, solutions based on a common intermediate structure could be a key supporting element. This support could be based on metrics, testing and deeper static analysis. Development of such support would introduce new values into software engineering field. For all offered reasons, proposed universal tree could be an appropriate internal representation applicable toward all stated goals. Universality of internal structure is important for meeting consistency in all fields.

By realization of this idea key benefit could be made from language independency of eCST and its universality and broad applicability.

### 4.1 Motivation

Motivation for introducing eCST as a new intermediate representation of the source code lays in intention to fulfil gaps in field of systematic application of software metrics by improving characteristics of software metric tools. One of the important weaknesses of available metric tools is the lack of support for calculation of metric values independently on input programming language.

Originally, Concrete Syntax Tree (CST) is used for representation of a source code. CST represents concrete source code elements attached to corresponding construction in language syntax. Although this tree is quite rich, it is still unaware of sophisticated details about meaning of syntax elements and their role in certain problems (e.g. Algorithms for calculation of software metrics). We enrich CST by adding universal nodes to mark elements to become recognizable independently on input programming language. To illustrate the technique to achieve independency of programming language we provide a simple example. It illustrates problems in calculation of Cyclomatic Complexity (CC) metric by predicate counting method.

The simple loop statement written in Modula-2 is stated as follows:
    REPEAT
        …Some statements…
    UNTIL (i > j);

The equivalent loop in Java would look like:
    do{
        …Some statements…
    }while (i <= j);

Although given statements have different syntax they express the same functionality: "some statements" in the code will be repeated until parameter "i" becomes greater then the parameter "j". Beside the different syntax, condition for leaving the loop is oppositely stated. First condition express what condition should be fulfilled to leave the loop, while the second one states condition to continue looping. Simplified syntax trees representing given statements are illustrated by Figure 1.

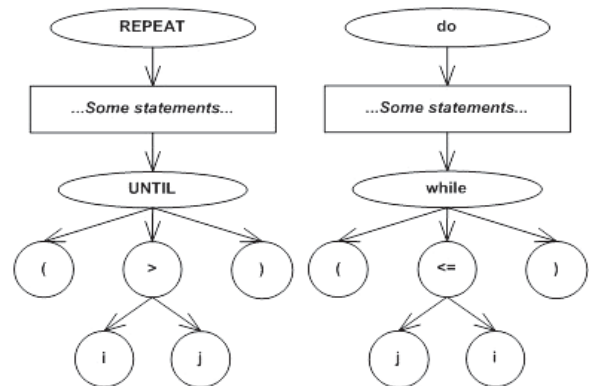

Figure 1: *Simplified syntax trees for REAPEAT-UNTIL (left) and do-while (right) statements*

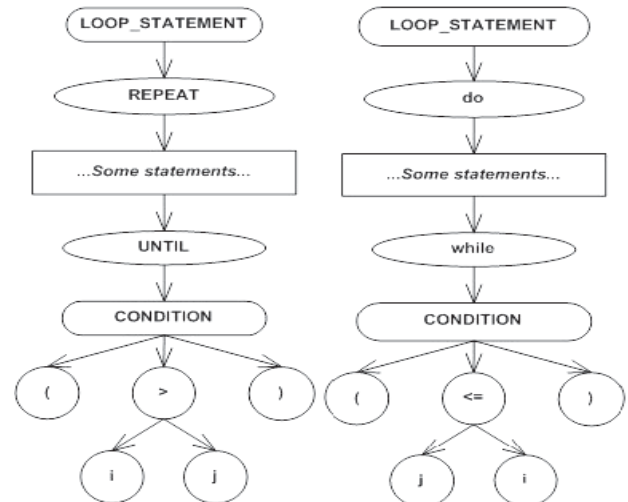

Figure 2: *eCSTs for REAPEAT-UNTIL (left) and do-while (right) statements*

For the implementation of CC algorithm we should recognize "REPEAT" and "while" as loops and to increment the current CC value by 1. It is clear that by using CST to represent source code we would need two implementations or at least two conditions to recognize these loops in the tree. By adding universal nodes "LOOP_STATEMENT" as parent of sub-trees that represent these two segments of source code we meet our goal by only one condition in implementation of the CC algorithm. Also we add universal node CONDITION to mark condition for leaving the loop repetition (Figure 2).

Additional enrichments for some other purposes could be including information about logical value that condition should have to leave the loop.

By adding all of needed universal nodes we implemented algorithms for CC metric independently on programming language. All we need is language grammar to modify and generate appropriate parser that is used for generating eCST.

New prototype of SMIILE tool that use eCST in metric calculation is described in [10]. This related paper describes language independent implementation of CC software metric based on universal nodes. It concentrates on CC as characteristic example for demonstration of usefulness of eCST in the direction of language independency of described tool. The paper provide table of used universal nodes in this prototype and provide way of usage in case of three characteristic languages – object-oriented Java, procedural Modula-2 and legacy COBOL.

### 4.2 Possible broader applicability

eCST was originally used in the development of language independent software metrics tool (SMIILE) [9]. Current prototype is implemented to support two software metrics and three languages.

However eCST has a limitation - it represents only separate compilation units. By translating all compilation units we get a set of autonomous trees. For the implementation of e.g. design metrics these trees should be connected by creating directed graph.

Idea for connecting compilation units is based on information about function (procedure, method, etc) calls contained in other functions. These calls could be placed either in the same or in some other compilation units.

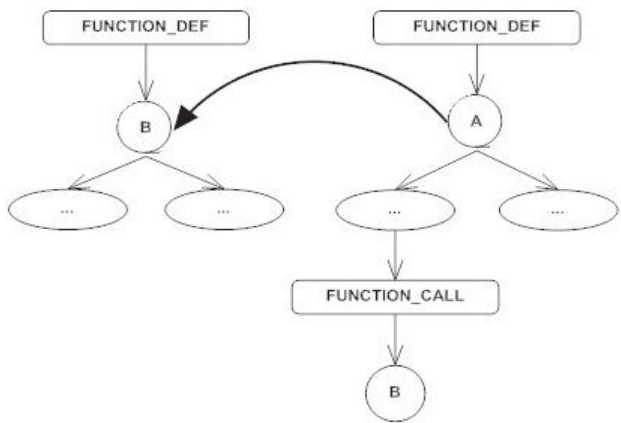

Figure 3: *Connecting of compilation units into call graph*

If function A contains call of function B than directed branch would lead from node representing function B to node that represent function A (Figure 3). Universal nodes (FUNCTION_DECL, FUNCTION_DEF and FUNCTION_CALL) would be used to locate the fragment of source code that contains function declaration, definition or call respectively. Information about function is placed in sub-tree of corresponding universal node.

Generated graph is a specific call graph. Maybe we can use even complex network (e.g. the one in [12]), but by creating the network by connecting eCSTs it would become language independent.

Additional possibility is to transform eCST to language independent enriched Control Flow Graph (eCFG), by inserting branches that represent possible execution paths through program (Figure 4).

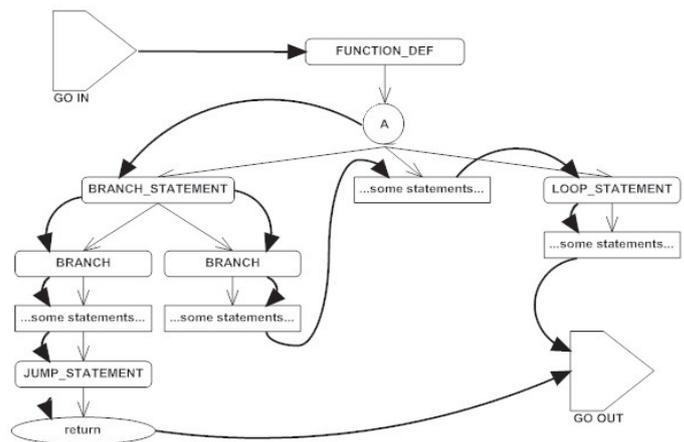

Figure 4: *Simplified view of eCST to eCFG transformation*

Generated eCFG could be used in software testing [4] (e.g. for development of automatic test case generator), in dead and duplicated code discovering, code-clone analysis [1],[2],[5], etc., but also as a basis for connecting compilation units instead of original eCST. In this case language independent call graph would be created by connecting eCFG components that represent compilation units.

Furthermore, we can notice that eCST could be used for automatic source code translation between programming languages. If we again consider given example we can conclude that given statement could be translated from Java to Modula-2 or from Modula-2 to Java. For automatic translation by using eCST we would have reflection table with rules for translation. In this example we should have rule about:

- How to translate the loop
- How to translate the condition
- How to translate inside statements

In this concept for translation we will not get perfectly written source code but by defining proper rules for translation we could manage to get correct source code. This limitation could be corrected by several cycles of code transformation [8].

SMIILE tool which is based on eCST for short-term goal had language independency, but as long term objective we stated smart software metrics tool that would recommend to developers how to improve their source code. It is planed to develop input language independent metric based advising system which would communicate with its user not only about metric values, but by concrete advices for corrections and refactoring of the source code based on calculated software metrics. For this purpose metric values should be stored in data storage. This storage could be well organized XML file system as primarily was proposed by SMIILE team, but also external software metrics repository could be used. Integration of SMIILE prototype with particular software metrics repository described in [11] is basis for further work in this direction.

Opportunity for improvement refactoring process gives additional value to described potential application of eCST

in code translation because needed after-translation refactoring could be automatically suggested or even applied.

The tool that integrates all described functionalities, including ones planed for SMIILE tool, would provide important features for consistent development of heterogonous software systems consisting of different components, implemented in different programming languages.

Furthermore SMIILE tool has possibility of keeping history of source code and corresponding software metrics. For keeping history of the source code eCST is stored to XML file created according to structure of eCST. By adding code-change analysis to the planed it would become important support in software reengineering process [6].

## 5 CONCLUSION

In this paper we introduce eCST and propose its usage in source code and model representation in development of universal tool to support different software engineering techniques and processes.

Idea for introducing eCST is supported by example of successful development of the prototype of language independent software metrics tool.

As this paper provide still fresh idea, it is clear that there exist numerous open questions and further work in proposed directions is planned.

**ACKNOWLEDGMENTS**

The authors acknowledge the support of this work by the Serbian Ministry of Education and Science through project "Intelligent Techniques and Their Integration into Wide-Spectrum Decision Support," no. OI174023.